\documentclass[twocolumn,showpacs,preprintnumbers,nofootinbib,prd,superscriptaddress,groupedaddress,10pt]{revtex4-1}

\usepackage{graphicx,amssymb,amsmath,amsthm,amsfonts,epsfig,epsf}
\usepackage[linktocpage]{hyperref}
\usepackage[usenames]{color}
\usepackage{epstopdf}

\usepackage{aas_macros}
\usepackage{bm}
\usepackage{dcolumn}
\usepackage[latin1]{inputenc}
\usepackage{latexsym}
\usepackage{rotating}
\usepackage{hyperref}
\usepackage{color}
\usepackage{longtable}
\usepackage{enumerate}
\usepackage{tensor}
\usepackage{url}
\newcommand{\ben}{\begin{enumerate}}
\newcommand{\een}{\end{enumerate}}

\def\be{\begin{equation}}
\def\ee{\end{equation}}
\def\bea{\begin{eqnarray}}
\def\eea{\end{eqnarray}}
\newcommand{\beq}{\begin{eqnarray}}
\newcommand{\eeq}{\end{eqnarray}} 
\newcommand{\ba}{\begin{align}}
\newcommand{\ea}{\end{align}}

\begin{document}

\title{Collapse of massive fields in anti-de Sitter spacetime}

\author{
Hirotada Okawa$^{1,2}$
,
Jorge C. Lopes$^{3}$
,
Vitor Cardoso$^{3,4}$ 
}
\affiliation{${^1}$ Yukawa Institute for Theoretical Physics, Kyoto University, Kyoto, 606-8502, Japan}
\affiliation{${^2}$ Advanced Research Institute for Science \& Engineering,
Waseda University, 3-4-1 Okubo, Shinjuku, Tokyo 169-8555, Japan}
\affiliation{${^3}$ CENTRA, Departamento de F\'{\i}sica, Instituto Superior T\'ecnico -- IST, Universidade de Lisboa -- UL,
Avenida Rovisco Pais 1, 1049 Lisboa, Portugal}
\affiliation{${^4}$ Perimeter Institute for Theoretical Physics Waterloo, Ontario N2J 2W9, Canada}


\begin{abstract}
Gravitational collapse in asymptotically anti-de Sitter spacetime has a rich but poorly-understood structure.
There are strong indications that some families of initial data form ``bound'' states, which are regular everywhere,
while other families seem to always collapse to black holes. Here, we investigate the collapse of massive scalar fields in anti-de Sitter,
with enlarged freedom in the initial data setup, such as several distinct wavepackets, gravitationally interacting with each other.
Our results are fully consistent with previous findings in the literature: massive fields, which have a fully resonant spectra, 
collapse at (arbitrarily?) small amplitude for some classes of initial data, and form oscillating stars for others.
We find evidence that initial data consisting on several wavepackets may allow efficient exchange of energy between them, and delay the collapse substantially, or avoid it altogether. When the AdS boundary is artificially changed so that the spectrum is no longer resonant, cascading to higher frequencies may still be present. Finally, we comment on the asymptotically flat counterparts.
\end{abstract}



\maketitle

\section{Introduction}

Numerical evidence indicates that general classes of initial data in anti-de Sitter (AdS) spacetime collapse to black holes (BHs), 
at very (perhaps even arbitrarily) small amplitudes~\cite{Bizon:2011gg}. One crucial ingredient for this phenomena is the confinement property
of the timelike boundary of global AdS. Although not rigorously proven, this is thought to be one necessary condition for collapse at arbitrarily small amplitudes. As some of us pointed out, the results in AdS geometries raise the intriguing possibility that analogous effects might play a role in our universe, by triggering collapse of compact stars, for example~\cite{Okawa:2014nea}.
In fact, similar characteristics -- collapse at very small amplitudes -- were observed in a ``box'' in flat spacetime under different boundary conditions~\cite{Maliborski:2012gx,Okawa:2014nea}.

By using a perturbative expansion of the field equations, 
the original study~\cite{Bizon:2011gg} suggested a second important
ingredient: that a fully-resonant --~or \emph{commensurable}~-- spectrum resulting in a secular term
is a necessary condition for collapse at {\it arbitrarily small
amplitudes}. Commensurability is a condition on the proper eigenfrequencies of the background spacetime.
In particular, there exist combinations of $i,j,k,l\in \mathbb{Z}$
to drive arbitrarily-small initial pulses away from the perturbative regime~\cite{Bizon:2011gg,Craps:2014vaa}
such that the corresponding modes satisfy
\beq
\omega_l=\omega_k +\omega_j-\omega_i\,,\quad l > i,j,k\,. 
\eeq
This condition guarantees that higher frequency modes are excited at full nonlinear level, potentially triggering a turbulent cascade and collapse. Such cascade to higher frequencies was indeed observed in the original and subsequent works~\cite{Bizon:2011gg}.
Nevertheless, because commensurability is generically violated at nonlinear level, it would imply -- should it be a necessary condition
for collapse -- that generic families of initial data do not collapse at small amplitudes. In line with this reasoning, there is indeed strong evidence that other classes of initial data do not collapse, forming instead bound states or stars~\cite{Buchel:2013uba,Dias:2012tq,Balasubramanian:2014cja}. 

In an attempt to understand the importance of ``turbulent'' gravitational behavior in asymptotically flat spacetime, 
some of us found a surprising behavior. In a toy model describing a scalar field confined in flat spacetime by a thin shell of matter, we found that collapse ensued at very small amplitudes and for a variety of boundary conditions.
For generic boundary conditions, the commensurability property is not satisfied, there were no hints of systematic cascading to higher frequencies,
but we still found consistent approach to gravitational collapse, something at odds with the current understanding~\cite{Okawa:2014nea}.

The numerical implementation used in Ref.~\cite{Okawa:2014nea} differs from those of other works on the subject,
and so does the specific setup. As such, our purpose here is to investigate collapse of scalar fields in AdS, 
using similar numerical procedures to those reported in the literature, but allowing more freedom in the system
so as to understand whether the conclusions in Ref.~\cite{Okawa:2014nea} were particular to that setup.
Our purpose is to explore the gravitational collapse of scalars in AdS, but allowing some more freedom in the theory and in the setup, in particular

\noindent (i) the scalar is generically massive,

\noindent (ii) the outer boundary may be located at a prescribed position, rather than only at spatial infinity (following Ref.~\cite{Buchel:2013uba}), and

\noindent (iii) The initial data will consist, generically, on a superposition of gaussian-like wavepackets. 

Condition (i) still implies a fully resonant spectrum for the scalar fluctuations, while condition (ii) breaks it. 

\section{Setup}

The system we are studying is a generalization of a system studied by other authors \cite{Bizon:2011gg}.
In our work we will use $c=\hbar=G=1$.
The model considered in this study describes the dynamics of a spherically symmetric massive real scalar field in a 3+1 dimensional AdS spacetime.
In such case the equations that rule the dynamics of the system are the Einstein equation and the Klein-Gordon equation, with a mass term,
\begin{eqnarray}
\resizebox{.85\hsize}{!}{$G_{ab} + \Lambda g_{ab} = 8\pi\left(\nabla_a\phi\nabla_b\phi - \frac{1}{2}g_{ab} (\nabla\phi)^2-g_{ab}V(\phi)\right)\label{eq:E_eq}$},\\
\frac{1}{\sqrt{-g}}\partial_{a}\sqrt{-g}g^{ab}\partial_{b}\phi - \frac{\partial V}{\partial\phi}=0\label{eq:KG_eq}.
\end{eqnarray}

We want to solve these equations for massive fields. As such we need to define what is their potential. We will set the massive potential to $V(\phi) = \frac{1}{2} \mu^2 \phi^2$.

In our study we will use as ansatz the line element
\begin{equation}
ds^2 = \frac{l^2}{\cos^2 x}(A e^{-2\delta} dt^2 +A^{-1}dx^2+\sin^2 x\ d\Omega^2)\label{eq:Met_Ansatz},
\end{equation}
where $l^2 = -3/\Lambda$ and $d\Omega^2$ is the differential solid angle of the unit two-sphere. We can recover the radial coordinate in Schwarzschild coordinates $r = l \tan x$.
The symmetric nature of the system demands that its variables, $A$, $\delta$ and $\phi$, depend only on the time and radial coordinates, $t$ and $x$.

Using the metric ansatz above together with the auxiliary variables $\Phi = \phi'$ and $\Pi = A^{-1}e^\delta\dot{\phi}$,
the Klein-Gordon equation \eqref{eq:KG_eq} yields
\begin{align}
 \dot\Phi &= \left(A e^{-\delta}\Pi\right)',\label{eq:Phi}\\
 \dot\Pi &=  \frac{1}{\tan^2 x}\left(\tan^2x A e^{-\delta}\Phi\right)' -\frac{l^2 \mu^2\phi}{e^{\delta}\cos^2x}.\label{eq:Pi}
\end{align}
The Einstein equations yield also
\begin{align}
 A' =& \frac{2(1+2\sin^2x)}{\sin 2x}(1-A)\nonumber\\
 &-4\pi\sin x\cos x A\left(\Phi^2+\Pi^2 \right)
 -4\pi l^2 \mu^2\phi^2\tan x,\label{eq:B}\\
 \delta' =& -4\pi\sin x\cos x\left(\Phi^2+\Pi^2\right)\,.\label{eq:delta}
\end{align}
By appropriately re-defining $\mu$ one can re-scale the AdS radius $l$
out of the problem. In other words, our results scale completely with $l$. 
When $\mu \to 0$ one recovers previous system of equations~\cite{Bizon:2011gg}.

For the solution to be physically meaningful, the profiles must be smooth everywhere. 
At the origin we set the boundary conditions
\begin{eqnarray}
 \phi(t,x) &=& \phi_0(t) +\mathcal{O}(x^2),\\
 A(t,x) &=& 1 +\mathcal{O}(x^2),\\
 \delta(t,x) &=& 0 +\mathcal{O}(x^2)\,,
\end{eqnarray}
and, at spatial infinity, we require that
\begin{eqnarray}
\phi(t,\rho) &\simeq& \phi_{\infty}(t)\rho^{k} + \mathcal{O}(\rho^{k+2}),\\
A(t,\rho) &\simeq& 1 -M\rho^3 +\mathcal{O}(\rho^5),\\
\delta(t,\rho) &\simeq& \delta_{\infty}(t) +\mathcal{O}(\rho^{2k}),
\end{eqnarray}
where $\rho\equiv\frac{\pi}{2}-x$ and $k = \frac{3}{2} +\sqrt{\frac{9}{4}+\mu^2\,l^2}$.
The behavior for the scalar field is dictated by regularity of the Klein-Gordon equation
in an asymptotically AdS spacetime.
\subsection{The eigenfrequencies}
%

\begin{table}[b]
\centering \caption{Resonant frequecies for massive scalars in AdS, when
 the boundary is truncted at a finite $r_{\rm max}/l=\tan{x_{\rm
 max}}$.} \vskip 12pt
\begin{tabular}{@{}c|cc@{}}
\hline \hline
$x_{\rm max}/l$&\multicolumn{2}{c}{$\omega\,l$}\\ \hline
                 &$\mu\,l=0$                  &$\mu\,l=2$  \\
                 &$3.0000$            &$4.0000$         \\
$\pi/2$          &$5.0000$            &$6.0000$         \\
                 &$7.0000$            &$8.0000$         \\ \hline
                 &$3.6497$            &$4.4463$        \\
$0.75$           &$7.1619$            &$7.7302$         \\
                 &$11.056$            &$11.471$        \\ \hline
                 &$9.3352$            &$10.198$        \\
$0.25$           &$20.452$            &$21.145$        \\
                 &$32.455$            &$32.964$         \\
\hline \hline
\end{tabular}
\label{ringtable1}
\end{table}  
To compute the eigenfrequencies of scalar fields in AdS at linearized level, it proves easier to work with AdS in (re-scaled, with the AdS radius $l=1$) global coordinates, 
\be
ds^2=-(r^2+1)dt^2+\frac{dr^2}{r^2+1}+r^2d\Omega^2\,.
\ee
The spherically symmetric, linearized Klein-Gordon equation for a field $\phi=\Psi/r\,e^{-i\omega t}$ reads
\be
f^2\frac{d^2\Psi}{dr^2}+ff'\frac{d\Psi}{dr}+\left(\omega^2-f\left(2+\mu^2\right)\right)=0\,,
\ee
with $f=r^2+1$. The above equation is of hypergeometric type. The solution which vanishes at the origin is
\be
\Psi=r(r^2+1)^{-\frac{\omega}{2}}F\left(\frac{3-k}{2}-\frac{\omega}{2},\frac{k}{2}-\frac{\omega}{2},\frac{3}{2},-r^2\right)\,,
\ee
Imposing Dirichlet conditions at infinity one finds
\be
\omega\,l=k+2n\,,
\ee
so that the spectrum is still fully resonant in the terminology introduced in the Introduction, and will give rise to
secular terms~\cite{Evnin:2015gma}.
One can break the resonant spectrum by imposing Dirichlet conditions instead at a finite radius, as done in Ref.~\cite{Buchel:2013uba}.
Examples of resonant frequencies are shown in Table~\ref{ringtable1}.

\subsection{Initial data}
%
In this paper, we use two types of initial data.
One is the single wavepacket initial data used in the literature~\cite{Bizon:2011gg},
\begin{align}
 \Phi(0,x)= 0,\quad
 \Pi(0,x) = \frac{2\epsilon}{\pi} e^{\frac{-4\tan^2 x}{\pi^2\sigma^2}}\,,
\end{align}
where $\sigma$ represents the width and $\epsilon$ the wavepacket's amplitude.
The other is composed of several wavepackets,
\begin{align}
 \Pi(0,x) = \epsilon\sum_{i=1}^{N}\frac{2a_{i}}{\pi} \exp\left[\frac{-4(\tan x-r_i)^2}{\pi^2\sigma_i^2}\right]\,,\label{init_several}
\end{align}
where we still assume $\Phi(0,x)= 0$ and, the width and location of the
$i-th$ wavepacket are defined by $\sigma_i, r_i$.
We also define overall amplitude by $\epsilon$ and the ratio of
amplitude to the first wavepacket by $a_i$, with $a_1=1$.

\section{Numerical results}
A description of the methods used to solve the system in the massless case are available in the literature~\cite{maliborski2013lecture}.
We generalized the approach, which uses a RK4 time evolution method with finite differences to compute the spatial profiles at each step of the evolution.
Note that it is possile to identify the apparent horizon (AH) formation
at $x_{AH}$ corresponding to a zero of $A(t,x)$ or a point at which
$A(t,x_{AH})$ falls below a prescribed
threshold\cite{Maliborski:2013via}.
Generically, our results show second-order convergence for the evolution of massive fields (and fourth-order for massless fields). A convergence test
is shown in Appendix~\ref{sec:convergence_test}.
\subsection{Massive field collapse}
%
\begin{figure}[ht]
 \includegraphics[width=80mm,clip]{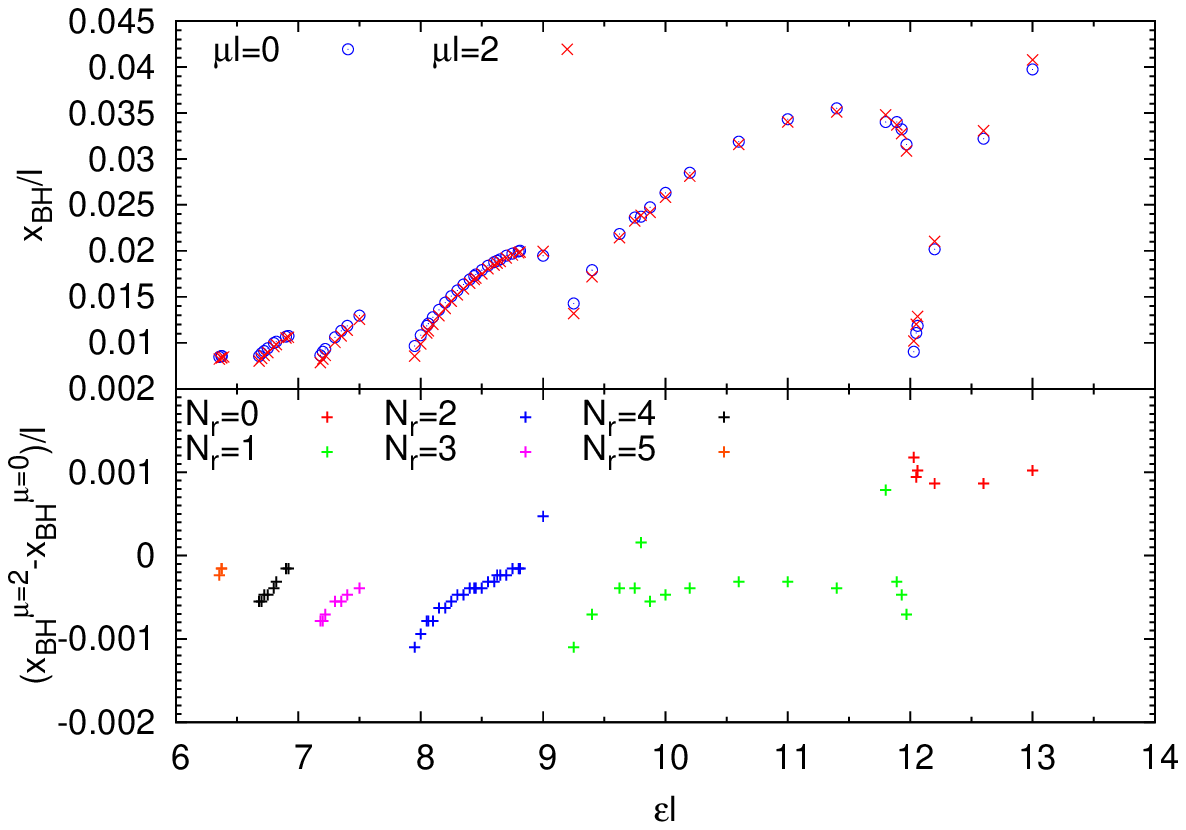}
 \includegraphics[width=80mm,clip]{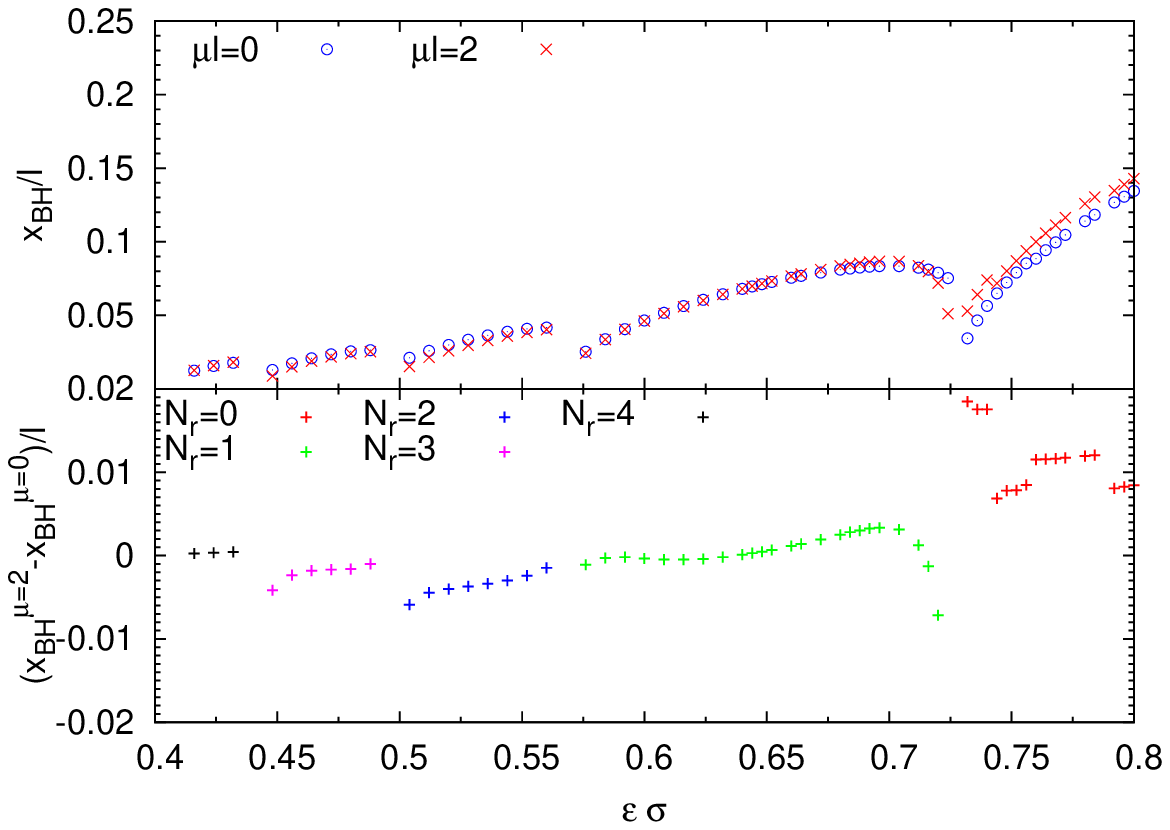}
 \caption{Direct comparison by BH radii formed through
 massless collapse($\mu l=0$) and massive collapse($\mu l=2$).
 The width/amplitude of the Gaussian wavepacket is fixed at $\sigma=l/16$ (top) and $\epsilon l=4$ (bottom).}
 \label{fig:comparison_mass}
\end{figure}
Our results for the gravitational collapse of a {\it single wavepacket} of massive fields
are best summarized by Fig.~\ref{fig:comparison_mass}.
Here, we compare directly the BH radius formed through the collapse
of a massless($\mu l=0$) and a massive($\mu l=2$) scalar for a
relatively small width $\sigma=l/16$ (top plot) and large amplitude $\epsilon l=4$ (bottom plot). For these, and
despite the relatively large value for $\mu$ -- larger than any other
scale in the problem -- collapse proceeds almost blindly with respect to the field mass.
Although not shown here, other mass terms (we tested for example $\mu l=1$) also exhibit the same trend.
The time elapsed until collapse also shows a remarkable agreement
for massive and massless fields. There are some fine prints, visible in the lower panels of the plots.
For fixed width and large initial amplitudes (within the ``Choptuik'' stage of prompt collapse), the mass term tends to make the BH size larger.
However, the BH seems to become smaller almost everywhere else at smaller amplitudes, with the exception of the critical points.

\begin{figure}[ht!]
 \begin{tabular}{c}
  \psfig{file=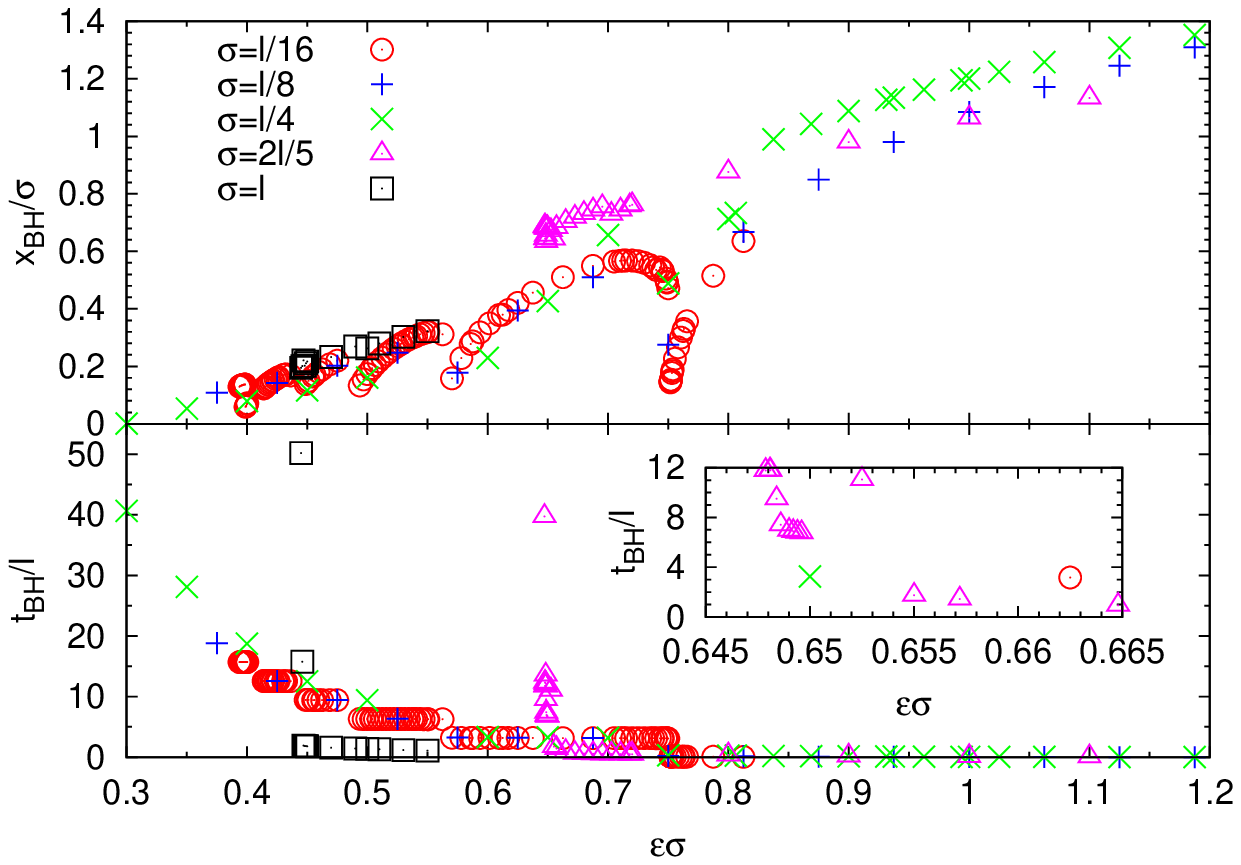,width=8.cm}\\
  \psfig{file=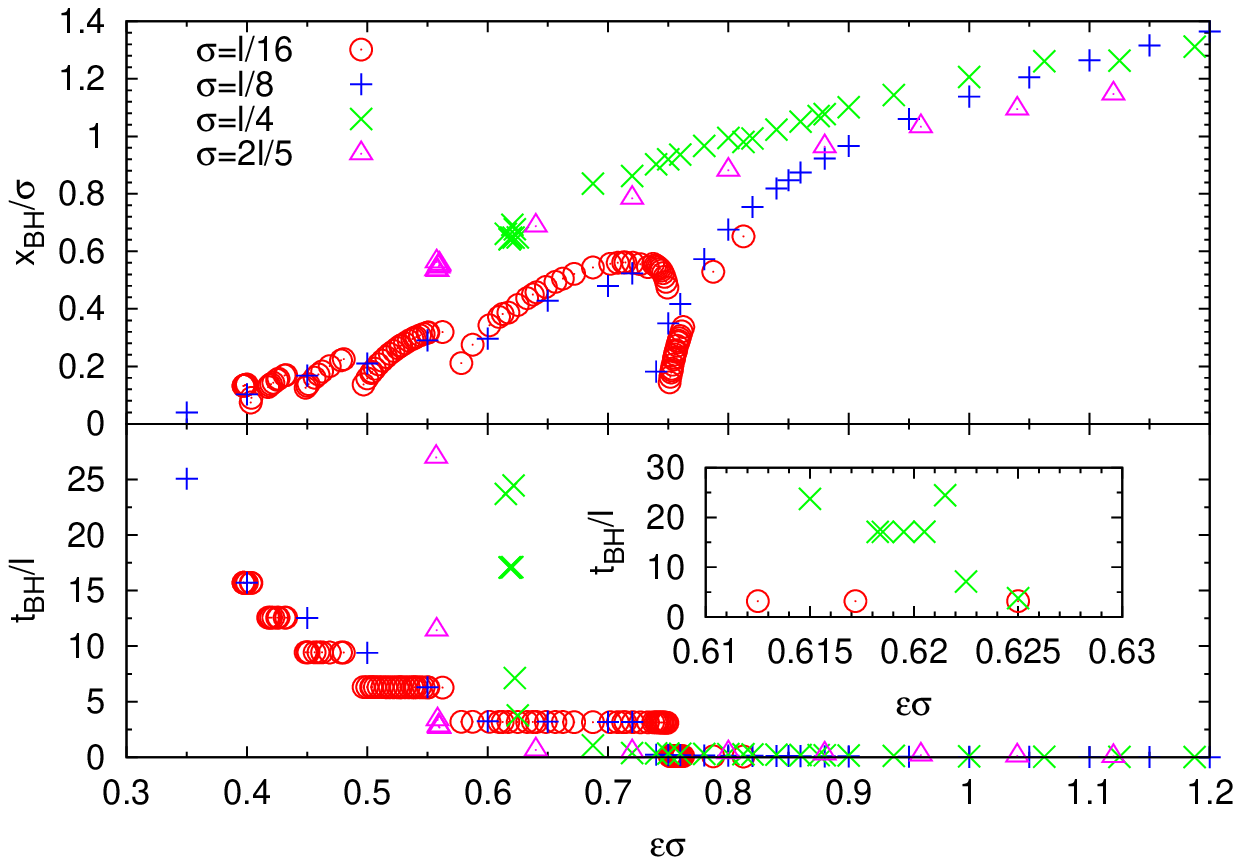,width=8.cm}
 \end{tabular}
 \caption[]{
 Critical behavior in AdS with different initial pulse width $\sigma=l/16,l/8,l/4$ and $2l/5$.
 Upper panels denote the BH radius while lower panels show the collapse time as a function of initial amplitude.
 Top: massless fields, $\mu l=0$. Bottom: massive fields, $\mu l=2$.
 }
 \label{fig:width}
\end{figure}
%

\subsection{Stability islands}
%
\begin{figure}[ht]
 \includegraphics[width=80mm,clip]{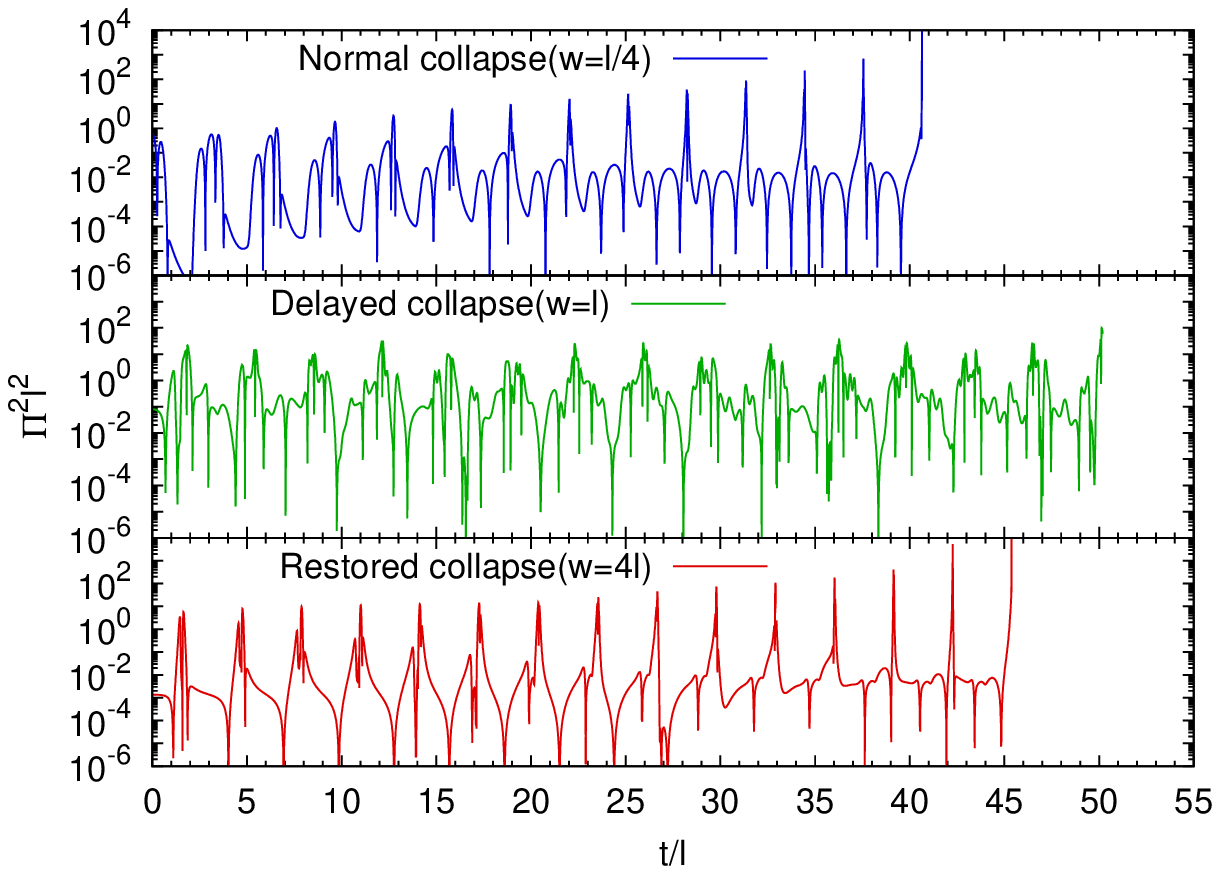}
 \includegraphics[width=80mm,clip]{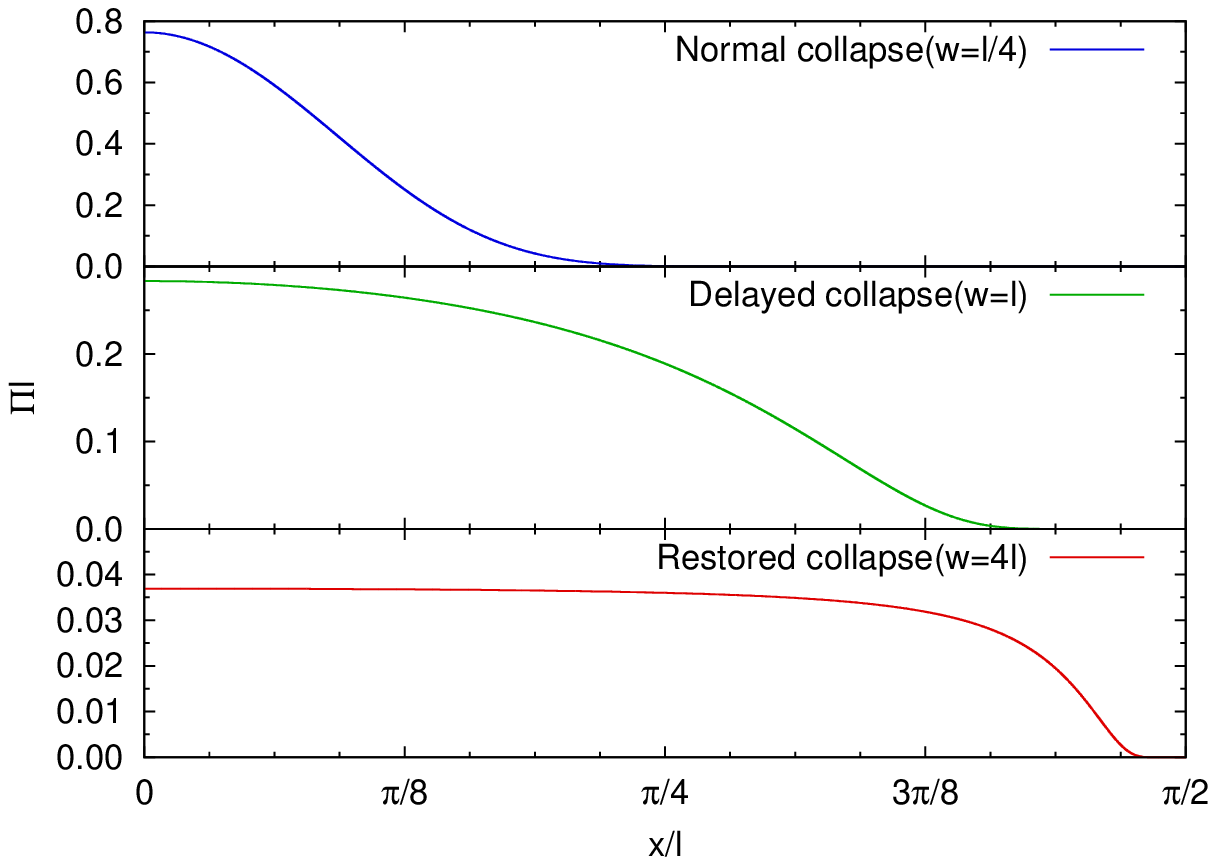}
 \caption{Top Panels: Ricci scalar evolution at the origin for ``normal'' (top),
 ``delayed'' (middle) and ``restored'' collapse (bottom) of massless fields.
 For very large widths, we find that collapse to BHs ensues again and that energy cascade to higher frequencies occurs efficiently.
 Bottom Panels: Corresponding initial profiles of the scalar field $\Pi$.
 }
 \label{fig:restored_collapse}
\end{figure}
\begin{figure}[ht]
 \includegraphics[width=80mm,clip]{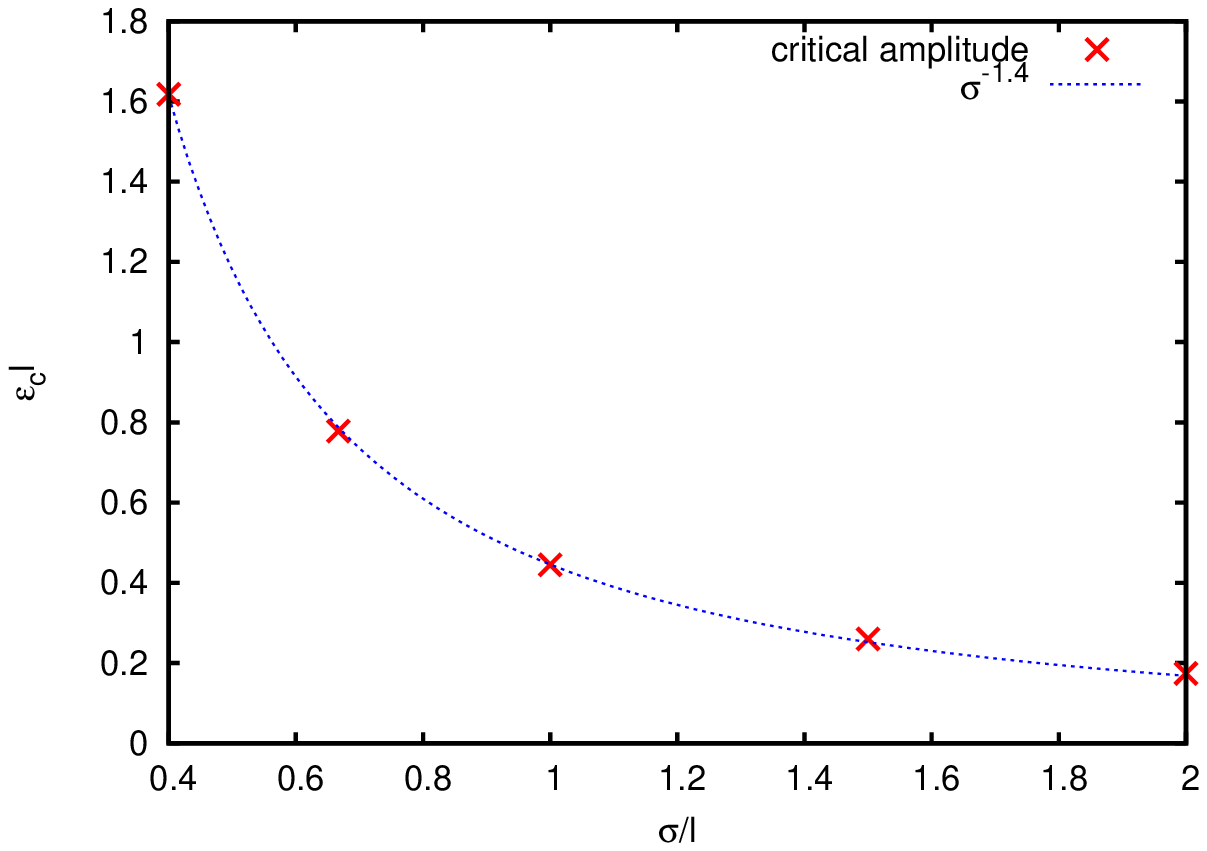}
 \includegraphics[width=80mm,clip]{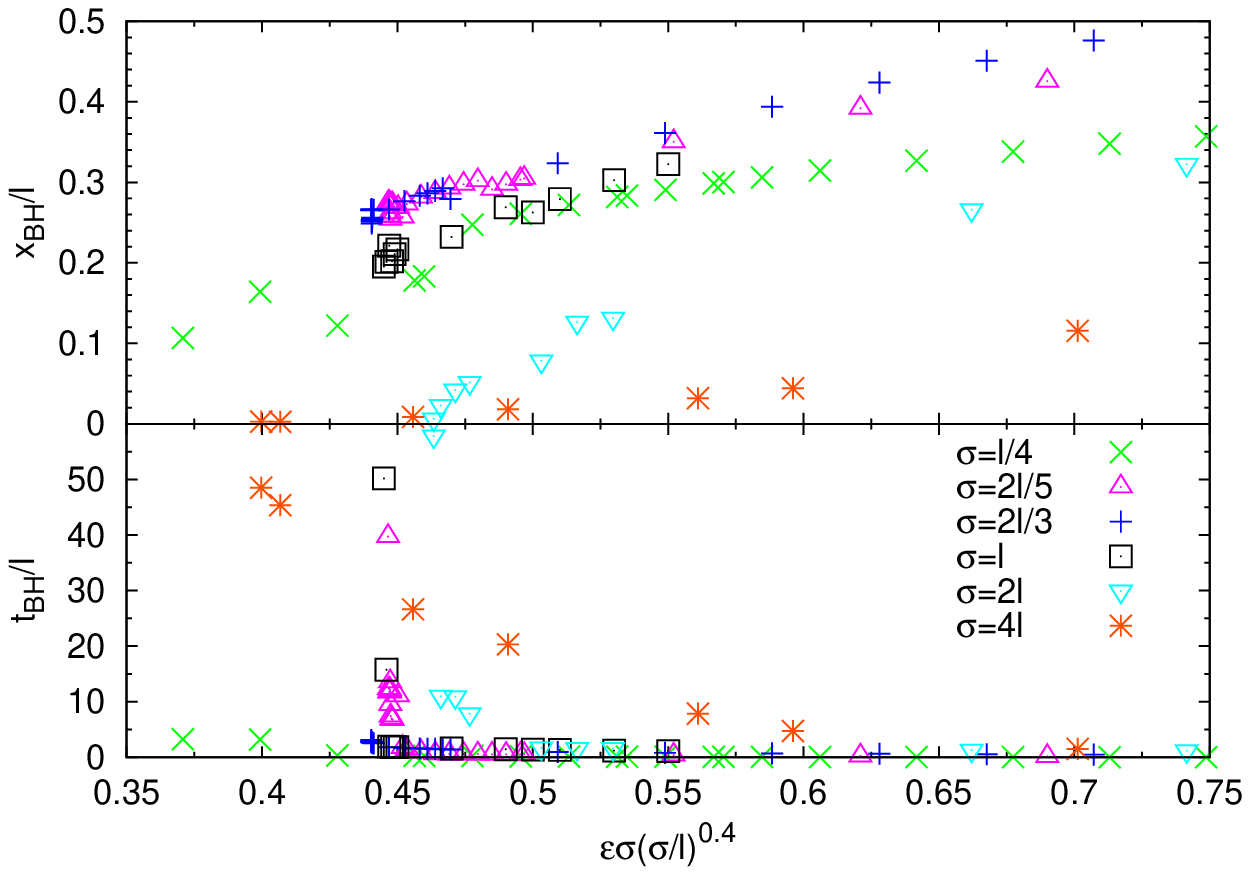}
 \caption{
 Delayed collapse of massless fields: Critical amplitude for collapse as a function of width (top).
 The bottom panels show the collapse time and BH size for re-scaled
 widths, where the contrast between small ($\sigma=l/4$), intermediate ($\sigma=l$) and very-large ($\sigma=4l$) initial width is apparent.}
 \label{fig:large_sigma}
\end{figure}

Stability ``islands'' corresponding to bound states, or boson stars in AdS, where no gravitational collapse is observed, were previously reported for massless fields~\cite{Buchel:2013uba}.
To investigate these possible end-states we have varied the width of the initial scalar profile, and monitored the collapse time.
The results are summarized in Figs.~\ref{fig:width}-\ref{fig:large_sigma} for massless ($\mu l=0$, upper panels) and massive ($\mu l=2$, lower panels) fields.
The results for massless and massive fields differ quantitatively, but not qualitatively.
In particular, we still find strong indications that there is no
collapse for families of initial data with moderately large width.
For example, $\sigma=2l/5$ takes an extremely long time to collapse below an amplitude $\epsilon\sigma\sim 0.65$;
on the other hand for a massive field with $\mu l=2$, not only $\sigma=2l/5$
but also $\sigma=l/4$ show extremely delayed collapse below $\epsilon\sigma\sim 0.62$.
There is thus a transition to new nonlinear solutions of the field equations;
these are known as oscillatons in asymptotically flat space~\cite{Seidel:1991zh} and have been observed to form in the collapse
of {\it massive} real fields~\cite{Okawa:2013jba}. In asymptotically AdS geometries oscillatons or ``scalar field breathers'' exist even for massless fields~\cite{Fodor:2013lza} and our results show that they are generically approached for some families of initial data.
The approach of collapsing configurations to oscillatons is not completely trivial as some of these oscillate for a long time before collapsing, presumably because they are either linearly or nonlinearly unstable~\cite{Okawa:2013jba}. This translates into a nontrivial collapse time, as the amplitude decreases, visible in the inset of Fig.~\ref{fig:width}.

It should also be noted that one can find whether such bound-states
arise or not by decreasing the initial amplitude and monitoring the final BH size near the critical amplitude, as in Fig.~\ref{fig:width}.
Our results imply that the transition from BHs to boson stars or bound-states is similar to that reported for the collapse of massive fields in a Minkowski background~\cite{Brady:1997fj,Okawa:2013jba}: the BH near the critical amplitude has a finite mass.
The confining \emph{box size} is now the AdS boundary,
while in previous Minkowski-background studies, it was simply the wavepacket width.

Moderately large-width initial data does not seem to collapse, and bound-states arise. 
As pointed out in Ref.~\cite{Maliborski:2013ula}, much larger widths actually restore the collapse, this is apparent in Fig.~\ref{fig:restored_collapse}, for massless fields.
Corresponding initial data are also shown in the lower panel. Although we do not show a detailed mode analysis, 
the transfer of energy to higher-frequency modes for ``normal'' (i.e., as reported originally in Ref.~\cite{Bizon:2011gg}) ($\sigma=l/4$) and ``restored'' ($\sigma=4l$) collapse, is already clear from the evolution of the Ricci scalar time evolution at the origin.
By contrast, the ``delayed collapse'' configuration shows nontrivial oscillation patterns.
We note that the initial data for restored collapse can be regarded as
a localized wavepacket with respect to the corresponding density distribution
shown in Ref.~\cite{Abajo-Arrastia:2014fma}.

For a region where delayed collapse is found,
we fitted the critical amplitude $\epsilon_c$ which divides bound-states from BHs in Fig.~\ref{fig:large_sigma}.
We find that $\epsilon_c\sim \sigma^{-1.4}$. Finally, the bottom panel shows that in a re-scaled axis plot, restored collapse ($\sigma=4l$) exhibits a completely different behavior from ``normal'' ($\sigma=l/4$) or delayed collapse($\sigma=l$).

\subsection{Finite boundary and resonant frequencies}
%
\begin{figure}[ht]
\includegraphics[width=80mm,clip]{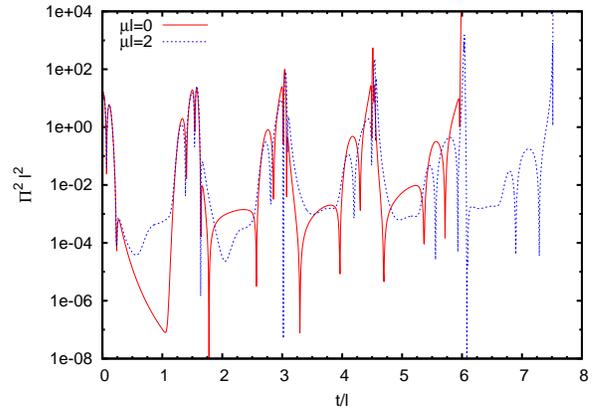}
\caption{Time evolution of the Ricci scalar $\Pi^2$ at the origin
during massless(red solid line) and massive(blue dotted line) collapse with an artificial boundary at $x_{\rm max}/l=0.75$.
Amplitude and width of the initial pulse are set by $\epsilon l=6.5$ and $\sigma=l/16$, respectively.
}
\label{fig:ricci}
\end{figure}
\begin{figure}[ht]
\includegraphics[width=80mm,clip]{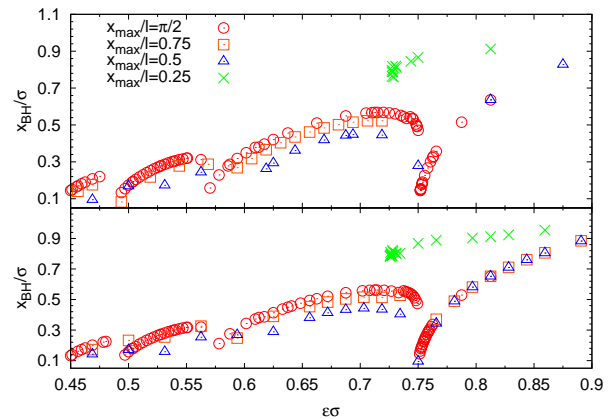}
\caption{Critical collapse with a finite boundary
for massless(top, $\mu l=0$) and massive (bottom, $\mu l=2$) scalars.
The outer boundary is located at a prescribed point,
by introducing an ``artificial mirror'' at $x_{max}$.
The width of scalar wavepackets is set by $\sigma=l/16$.
For $x_{max}/l=\pi/2$ one recovers Fig.~\ref{fig:width}.
}
\label{fig:stability_islands}
\end{figure}
\begin{figure}[ht]
 \includegraphics[width=80mm]{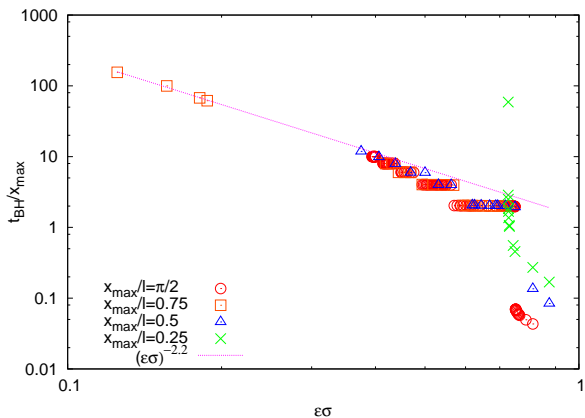}
 \caption{Rescaled collapse time for massless fields with a width $\sigma=l/16$.
 Different marks correspond to different mirror locations as
 Fig.~\ref{fig:stability_islands}.
 Dotted line was obtained by fitting the results from $x_{max}/l=0.75$ simulations.
 }
 \label{fig:collapse_time}
\end{figure}

The behavior reported previously, where some classes of (small-width) initial data collapse at very small amplitudes whereas others
(larger-width) do not, is not dependent on the resonant spectrum of AdS.
This can be tested by introducing an artificial ``mirror'' in the AdS bulk, which amounts to imposing Neumann boundary conditions for $\phi$ at that location.
The collapse properties are summarized in Figs.~\ref{fig:ricci}-\ref{fig:collapse_time}.

In Fig.~\ref{fig:ricci}, we show time evolutions of the Ricci scalar at the origin
for a massless($\mu l=0$) and massive($\mu l=2$) collapse with an artificial mirror boundary at $x_{max}/l=0.75$.
Notice how the scalar curvature increase super-exponentially for both cases, despite the fact that both clearly have a non-resonant spectra.

The structure of collapse time as function of amplitude and width shows an interesting structure depicted in Figs.~\ref{fig:stability_islands}-\ref{fig:collapse_time}.
For a wide range of initial conditions, the time development is not strongly dependent on a mass term, as is evident from Fig.~\ref{fig:stability_islands}.
We also show the rescaled collapse time with an artificial mirror at $x_{max}$ in Fig.~\ref{fig:collapse_time}.
Delayed collapse is clearly seen for $x_{max}/l=0.25$,
while the collapse time with finite boundary for a relatively large
$x_{max}$ also scales as $\epsilon^{-2}$, for example, $x_{max}/l=0.75$.

In summary, there are certainly families of initial data which do not collapse, in agreement with the view that backgrounds with non-resonant frequencies
cannot cause collapse at arbitrarily small frequencies. On the other hand, we do find collapse for other generic families of initial data {\it at finite}
amplitude (as must be the case numerically). In addition, we continue to observe migration to higher frequencies and a collapse time which is of the order of the full (resonant)
AdS spacetime. Unfortunately, our results do not shed new light on this issue, and the results in Refs.~\cite{Maliborski:2012gx,Okawa:2014nea} continue to be poorly understood (unless one admits that collapse will eventually
occur at a very small but finite amplitude, whose threshold is at the moment unpredicted).

\subsection{Energy transfer between wavepackets}
%
\begin{figure}[ht]
 \includegraphics[width=80mm,clip]{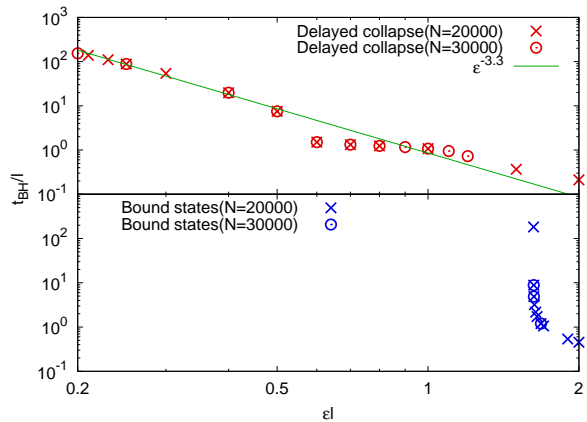}
 \caption{Collapse time for two different initial data sets: two (top)
 and three-wavepacket initial data (bottom).
 The collapse time is shown as a function of the overall amplitude
 $\epsilon l$ in \eqref{init_several}.
 Parameters are set by $\{\sigma_1,r_1,a_1\}=\{l/8,l \pi /8,1\}$ and
 $\{\sigma_2,r_2,a_2\}=\{l,2l\pi,1/10\}$ for two wavepackets and by
 $\{\sigma_1,r_1,a_1\}=\{l/4,0,1\}$, $\{\sigma_2,r_2,a_2\}=\{l/4,l\tan(\pi/8),1/4\}$
 and $\{\sigma_3,r_3,a_3\}=\{l/4,l\tan(\pi/4),1/8\}$ for three wavepackets, respectively.
 Crosses(Circles) denote the results with low(high) resolution.
 Notice that, although the three-wavepacket initial data does not collapse,
 the width of each individual wavepacket is very small, and would lead to collapse were they in isolation.
 }
 \label{fig:comparison_2and3solitons}
\end{figure}
\begin{figure}[ht]
 \includegraphics[width=80mm,clip]{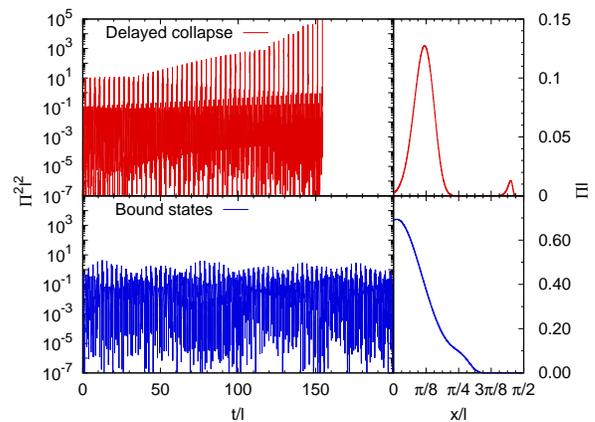}
 \caption{Ricci scalar evolution and corresponding initial data.
 Parameters are the same as in Fig.~\ref{fig:comparison_2and3solitons},
 and the overall amplitudes are given
 by $\epsilon l=1/5$ for two pulses and $\epsilon l=1$ for three pulses.
 In the top panel, collapse begins in the smaller-amplitude wavepacket.
 }
 \label{fig:Ricci_2and3solitons}
\end{figure}
\begin{figure}[ht]
 \includegraphics[width=80mm,clip]{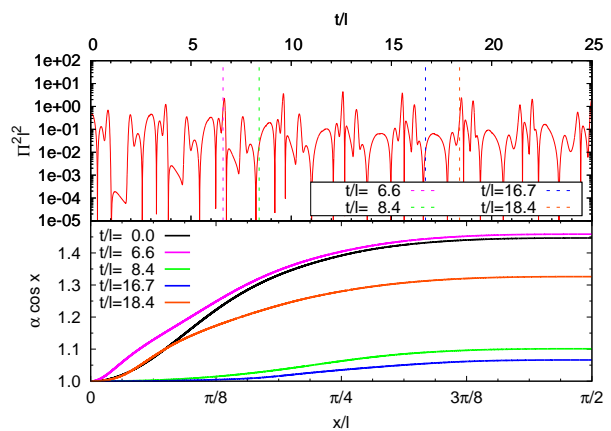}
 \caption{
 Early phase evolution of Ricci scalar at the origin(top) using three
 wavepackets initial data and snapshot of lapse function at $t/l=0.0, 8.4, 16.7, 24.5$ (bottom).
 Parameters are the same as those in Fig.~\ref{fig:Ricci_2and3solitons}.
 }
 \label{fig:3solitons}
\end{figure}

Our results do shed some more light on the existence of nonlinearly stable bound states, and the role of gravitational
redshift in their formation. For this purpose, we now focus on the collapse of initial data described by more than
one wavepacket.

In Fig.~\ref{fig:comparison_2and3solitons}, for example, we show the time it takes for collapse, starting from initial data described by two (upper panel)
and three wavepackets (lower panel). It is apparent from these plots that even relatively large amplitudes need some more reflections to collapse,
as compared with a single wavepacket.
In fact, the three-wavepacket configuration clearly displays delayed collapse, and exhibits a threshold amplitude below which no black hole is formed.
We should also add that the importance and effect of wavepacket superposition for gravitational collapse are also discussed in Ref.~\cite{Abajo-Arrastia:2014fma}.

Typical evolutions of Ricci scalar monitored at the origin are
exemplified in the left column of Fig.~\ref{fig:Ricci_2and3solitons} and corresponding initial data are shown in the right column.
The Ricci scalar grows monotonically for two wavepackets, while it oscillates almost periodically for three wavepackets. A mode analysis of
the three-wavepacket evolution shows a cascade to higher frequencies at early times, and a constant frequency content at late times.

In order to understand the finer structure of the evolution, 
the top panel of Fig.~\ref{fig:3solitons} shows only the early
phase evolution of Ricci scalar. There are some features worth highlighting.

\noindent (i) The initial (three-wavepacket) configuration corresponds roughly to the three peaks observed before $t/l\lesssim1$ and is similar
to that within $3<t/l<4$ (after one reflection) and $6<t/l<7$ (after two reflections).
However, the initial configuration produced, besides these in-going waves, out-going waves.
These are observed around $2<t/l<3$ or at around $t/l=5$ and correspond
to waves directed to the AdS boundary. Let us call them ``secondary waves''.

\noindent (ii) The original in-going packets (say at $t=0, 3, ~6, ~8.5, 11.5$) are always accompanied 
by high amplitude peaks to their right. The bottom panel of Fig.~\ref{fig:3solitons} shows that these are
in a higher gravitational potential, and can work as a source of redshift for those packets.
Conversely, these high-amplitude peaks can be blueshifted. In other words, these localized packets can become broader or sharper,
depending on the gravitational potential that they encounter.

\noindent (iii) However, we find that the secondary waves can ``migrate'' and even take-over the originally in-going pulses (in this example, after $t\sim 17$).
As a result, the redshift/blueshift pattern is inverted and processes continues in a cyclic way.

Since the scalar field is responsible for the spacetime curvature in this setup,
general-relativistic effects should become important when high-amplitude waves are located near the origin.
The bottom panel of Fig.~\ref{fig:3solitons} shows the lapse
function defined by $\displaystyle\alpha\equiv\sqrt{A}e^{-\delta}/\cos x$ in the line element~\eqref{eq:Met_Ansatz}.
What this really shows is how clocks tick differently as a function of coordinate $x$ for different instants, 
and can explain the migration of secondary waves as mentioned above, considering that high-amplitude
waves act as a source of gravitational potential and other waves apart farther from the origin move faster than the ones near the origin.
Also, the time dependency of lapse profile gives evolutions of original and secondary waves,
because the frequency of ingoing (outgoing) waves in the gravitational potential can be changed
by gravitational blueshift (redshift).
For a free-fall observer, the difference of frequencies between waves at
$x=x_1$ and $x=x_2$ is described formally by $\omega_2/\omega_1=\alpha (x_1)/\alpha (x_2)$
if the background spacetime is fixed.
That implies that higher-amplitude and wider-width waves could cause the
frequency shift much more efficiently than the ones slightly farther from the origin.
The existence of many such wavepackets would give a meta-stable state, continuously exchanging energy.

The above is, of course, a tempting explanation of why this type of initial data does not lead to collapse, but is far from rigorous.

\section{Discussion}
We investigated the collapse of spherically symmetric scalar fields in AdS for massive fields for generic initial data,
and including an artificial ``mirror,'' extending previous studies on
the subject. Our main motivation was to understand role of ``commensurability''
in the process and the mechanism leading to collapse or to stable configurations. Our main findings are that

I. In parallel with previous studies, we also find that there are
stability islands for some regions of parameters in the space of initial conditions.
In these regions -- both for massless and massive fields -- we observe quasi-stable bound states, or oscillating configurations.

II. Single-wavepacket initial data for moderately-large widths evolve into nonlinearly-stable configurations at very small amplitudes (see also Fig.~7 in Ref.~\cite{Buchel:2013uba}) and at finite amplitude it results in a delayed collapse, as summarized in Fig.~\ref{fig:large_sigma}.
However, we also confirm that much wider initial data restores the collapse again,
indicating that there are thresholds dividing the phase of nonlinear solutions into two final states.
Unlike turbulent collapse, the threhsold amplitude yields a finite-mass BH, in a behavior very similar to that found in the collapse of massive fields in asymptotically flat spacetimes~\cite{Brady:1997fj,Okawa:2013jba}.

III. Even if the condition of commensurability is violated, for example by inserting an artificial mirror at a finite distance, our results
exhibit energy transfer to higher modes and indicate the ``usual'' $t\sim\epsilon^{-2}$ dependence (see Fig.~\ref{fig:collapse_time}).
In addition, for smaller mirror radius, we again observe nonlinearly stable, oscillating bound states.
Our results suggest that the threshold of the initial pulse width for transitions is around $\sigma/x_{max}\sim 0.2$
for single wavepacket initial data. 

IV. We have argued that other, more generic type of initial data can produce stable bound states by an interplay of blueshift/redshift effects,
and that this mechanism may even be at play in large-width, single-wavepacket data. For example, data consisting on two wavepackets seem to 
efficiently transfer energy from one wavepacket to the other, significantly delaying their collapse as compared to the single-wavepacket data.
Interestingly, three-wavepacket initial data can evolve into stable bound-states, very much like moderately large-width initial data consisting on a single wavepacket.

We should highlight the fact that the collapse of massive field exhibits qualitatively the same behavior as that of massless field in our mass parameter choice; it is possible that much larger field mass produces more easily bound states, which would be the analog of flat-space oscillatons. As a final comment, we should stress that our findings, both in this work and in previous~\cite{Okawa:2014nea}, flat-space studies are perfectly consistent with the prediction that gravitational collapse at arbitrarily small amplitudes requires commensurable spectra~\cite{Dias:2012tq}. Non-commensurable spectra yields collapse at finite amplitudes, which might be of some physical consequence
in a variety of scenarios. An important open question concerns the threshold amplitude when the spectra is non-commensurable.
\appendix
\section{Convergence test}\label{sec:convergence_test}
\begin{figure}[ht]
 \includegraphics[width=85mm,clip]{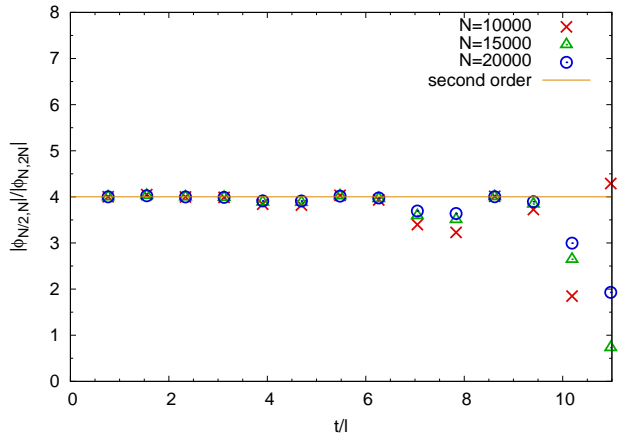}
 \caption{
 Convergence tests for the scalar field $\phi$ setting the parameters of
 initial pulse by $\sigma=l/16, \mu l=2$ and $\epsilon l=6.8$.
 The second order convergence with different resolutions $N_i=N/2,N,2N$ is shown by
 defining the L2-norm,
 $|\phi_{N_1,N_2}(t=t_n)|\equiv\left[\sum_{t=n\pi/2}^{(n+1)\pi/2}\left(\phi_{N_1}(t,0)-\phi_{N_2}(t,0)\right)^2\right]^{1/2}$.
 }
 \label{fig:convergence_tests}
\end{figure}

In Fig.~\ref{fig:convergence_tests}, we show convergence tests of the scalar field $\phi$ as a function of time
for massive collapse corresponding to Fig.~\ref{fig:comparison_mass}.
Our results are compatible with second order convergence for $\phi$ (obtained by a trapezoidal integration of $\Phi$, cf. main text above \eqref{eq:Phi}).
%

\begin{acknowledgments}
We are grateful to \'Oscar Dias, Oleg Evnin, Javier Mas and Jorge Santos for helpful discussion.
V.C. acknowledges financial support provided under the European
Union's FP7 ERC Starting Grant ``The dynamics of black holes: testing
the limits of Einstein's theory'' grant agreement no. DyBHo--256667,
and H2020 ERC Consolidator Grant ``Matter and strong-field gravity: New frontiers in Einstein's theory'' grant agreement no. MaGRaTh--646597.
This research was supported in part by the Perimeter Institute for
Theoretical Physics. Research at Perimeter Institute is supported by
the Government of Canada through Industry Canada and by the Province
of Ontario through the Ministry of Economic Development $\&$
Innovation.
This work was supported by the NRHEP 295189 FP7-PEOPLE-2011-IRSES
Grant.
\end{acknowledgments}

\vskip 5cm

\bibliographystyle{h-physrev4}
\bibliography{Turbulence_massive_refs}

\end{document}